\documentclass[twocolumn,showpacs,superscriptaddress]{revtex4}
\usepackage{amssymb}
\usepackage{amsmath}
\usepackage{graphicx}
\usepackage{epsfig}
\usepackage{amsfonts}

\begin{document}

\title{ Two Mode Photon Bunching Effect as Witness of Quantum Criticality in
Circuit QED }
\author{Qing Ai}
\affiliation{Department of Physics, Tsinghua University, Beijing 100084, China}
\author{Ying-Dan Wang}
\affiliation{Department of Physics, University of Basel, Klingelbergstrasse 82, 4056
Basel, Switzerland}
\author{Guilu Long}
\affiliation{Department of Physics, Tsinghua University, Beijing 100084, China}
\affiliation{Tsinghua National Laboratory for Information Science and Technology, Beijing
100084, China}
\author{C. P. Sun}
\affiliation{Institute of Theoretical Physics, Chinese Academy of Sciences, Beijing,
100080, China}

\begin{abstract}
We suggest a scheme to probe critical phenomena at a quantum phase
transition (QPT) using the quantum correlation of two photonic modes
simultaneously coupled to a critical system. As an experimentally
accessible physical implementation, a circuit QED system is formed
by a capacitively coupled Josephson junction qubit array interacting
with one superconducting transmission line resonator (TLR). It
realizes an Ising chain in the transverse field (ICTF) which
interacts with the two magnetic modes propagating in the TLR. We
demonstrate that in the vicinity of criticality the originally
independent fields tend to display photon bunching effects due to
their interaction with the ICTF. Thus, the occurrence of the QPT is
reflected by the quantum characteristics of the photonic fields.
\end{abstract}

\maketitle

\section{Introduction}

Inspired by the fast developments of quantum information
\cite{Osterloh,Gu}, quantum phase transition (QPT) \cite{Sachdev}
has renewed much attention in different fields of physics ranging
from condensed matter physics to quantum optics \cite{Emerly,Li}. It
was found that \cite{Quan06} at the quantum critical point the
dynamic evolution of a quantum critical system is so extremely
sensitive that it can enhance the quantum decoherence of an external
system coupled to it. This ultra-sensitivity is characterized by the
Loschmidt echo, which is a well-known concept in quantum chaos
\cite{Perese}. In this sense, the quantum-classical transition from
a pure state to a mixed one is induced by the quantum criticality of
this surrounding system. This discovery motivated a new scheme to
probe the QPT by exploring the quantum coherence in the external
system and its losses \cite{Wang07}.

Moreover, this probing mechanism for quantum criticality was
illustrated by a circuit QED architecture
\cite{Wallraff,Chiorescu,You}, which was formed by a superconducting
Josephson junction qubit array interacting with a one-dimensional
superconducting transmission line resonator (TLR) \cite{Wang07}. The
superconducting qubit array was modeled as an Ising chain in the
transverse field (ICTF). This investigation showed that the QPT
phenomenon in the superconducting qubit array was evidently revealed
by the correlation spectrum of TLR output. Though this mechanism for
the circuit QED system has not been experimentally tested, an NMR
simulation experiment \cite{Zhang} has been carried out to
demonstrate the QPT-like phenomenon (energy level crossing ) as
predicted in ref.\cite{Quan06} by exploring the increased
sensibility of the quantum system to perturbation when it is close
to a critical point.

For the above circuit QED architecture to demonstrate the probing of
the QPT, we notice that with two modes simultaneously coupled to a
charge qubit, their squeezing effect was investigated theoretically
\cite{Moon}. Here, we consider the full application of quantum
optics approach \cite{Scully} in the detection of the QPT by
considering the higher order quantum coherence. To this end, we
consider that a Josephson junction qubit array modeled as the ICTF
simultaneously couples to two modes propagating in the TLR. Since
all quasi-spins homogeneously interact with the fields, we can
obtain the first (second) order correlation function of the two
fields. According to our calculation, the second order quantum
coherence is given in terms of the decoherence factor of the ICTF.
As proven in the Appendix, the norm of the decoherence factor
decreases exceptionally when the ICTF is at the critical point.
Therefore, the photon bunching effect occurs since the second order
quantum coherence of the steady state is smaller compared with its
initial value. And these results show genetic characteristics of the
quantum spin chain in the vicinity of the critical point.

The paper is structured as follows. The next section describes the ICTF
formed by a capacitively coupled Josephson junction qubit array is coupled
to two independent fields propagating in the TLR. Then the detection scheme
of the QPT and the correlation functions of two mode fields are given in
Sec. III. A brief summary is concluded in Sec. IV. Furthermore, in addition
to the main body of the paper, Appendix \ref{app:appendix1} presents details
about the calculation of the decoherence factor.

\section{Circuit QED based Setup for Probing Quantun criticality}

We consider a circuit QED system illustrated in Fig.1. $N$ Cooper
pair boxes (CPBs) are capacitively coupled one by one. Formed by a
superconducting island connected with two Josephson junctions, each
CPB is a direct current superconducting quantum interference device
(dcSQUID). Since the magnetic flux $\Phi _{x}$ threading the dcSQUID
is tunable, the effective Josephson tunnelling energy can be varied.
With proper bias voltage, the CPB behaves as a qubit near the
degeneracy point and then Josephson junction qubit array becomes a
spin chain with $N$ 1/2-spins. When the coupling
capacitance $C_{m}$ between two CPBs is much smaller than the total one $%
C_{\Sigma }$ to each CPB, the high order terms in Hamiltonian can be
neglected and only the nearest neighbor interaction is considered.
Then the qubit array can be approximated as an ICTF with
\cite{Wang07}
\begin{equation}
H_{0}=B\sum\limits_{j=1}^{N}(\lambda \sigma _{j}^{x}+\sigma _{j}^{z}\sigma
_{j+1}^{z})\text{,}  \label{H0}
\end{equation}%
where $\sigma ^{x}=-|0\rangle \langle 1|-|1\rangle \langle 0|$ and $\sigma
^{z}=|0\rangle \langle 0|-|1\rangle \langle 1|$ with $|n\rangle $ being the
state of $n$ extra Cooper pair on the superconducting island, $\lambda
=B_{x}/B$ and $B=e^{2}C_{m}/C_{\Sigma }^{2}$, $B_{x}=E_{J}\cos (\pi \Phi
_{x}/\Phi _{0})$ is the Josephson energy of each CPB with $E_{J}$ the
Josephson energy of single junction, $\Phi _{0}=h/2e$ the flux quantum.

\begin{figure}[ptb]
\includegraphics[bb=115 314 485 516,width=7 cm]{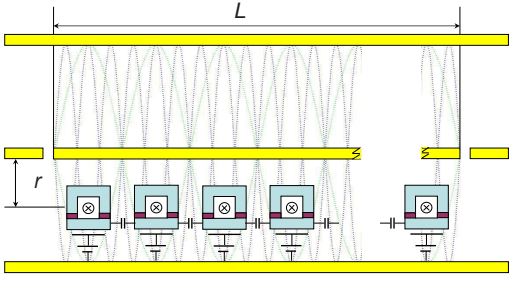}
\caption{ (color online) Schematic diagram of superconducting Ising
chain
interacting with two largely detuned modes ($\protect\omega_1=3\protect\omega%
_2$) individually transmitted in the TLR. Since the CPBs are located at the
antinodes of both modes, i.e, $x_j=(j-1/2)\protect\pi c/\protect\omega_2$,
they only interact with the magnetic fields. }
\label{Sketch}
\end{figure}

In a one dimensional TLR, the electric current and voltage at the position $x
$ are given as
\begin{align}
I(x,t)&=\sum_k\sqrt{\frac{\hbar\omega_k}{Ll}}(a_k^\dagger+a_k)\sin(\frac{%
k\pi x}{L}) \text{,} \\
V(x,t) &=-i\sum_k\sqrt{\frac{\hbar\omega_k}{Lc}}(a_k^\dagger-a_k)\cos(\frac{%
k\pi x}{L}) \text{,}
\end{align}
where $a_k^\dagger$ is the creation operator with frequency
$\omega_k$, $L$ the length of TLR, $l$ and $c$ the induction and
capacitance of per unit length of TLR respectively, $k$ positive
integer. Therefore, a CPB located at the antinode is only coupled to
the magnetic field since the electric field vanishes. According to
Ampere's circuital law, when a dc SQUID loop is placed at a distance
$r$ with respect to the center of the TLR, the quantum magnetic flux
that threads it is
\begin{align}
\Phi_q(x)&=\frac{\mu_0IS}{2\pi r}  \notag \\
&=\sum_k\frac{\mu_0S}{2\pi r}\sqrt{\frac{\hbar\omega_k}{Ll}}%
(a_k^\dagger+a_k)\sin(\frac{k\pi x}{L}) \text{,}
\end{align}
where $\mu_0$ is the vacuum magnetic permeability, $S$ the area of
dc SQUID loop.

The interaction between the CPBs and the magnetic field is written as
\begin{align}
H_\Phi=E_J\sum_j\cos(\frac{\pi\Phi_q(x_j)}{\Phi_0})\sigma^x_j \text{.}
\end{align}
In our consideration, two independent modes with frequencies $
\omega_1=3\omega_2$ are propagating in the TLR. All the CPBs are
placed the antinodes of the both modes with the positions
\begin{equation}
x_j=\frac{(j-1/2)\pi v}{\omega_2} \text{,}
\end{equation}
where
\begin{equation}
\omega_2=\frac{\pi vL}{M} \text{,}
\end{equation}
$j$ and $M$ are positive integers, $v$ the velocity of the light.

Since $\Phi _{q}\ll \Phi _{0}$, under the rotating wave approximation \cite%
{Scully}, the interaction Hamiltonian is approximated to the second order,
\begin{equation}
H_{\Phi }=E_{J}[(1-\frac{1}{2}\eta _{1}^{2}-\frac{1}{2}\eta _{2}^{2})-\eta
_{1}^{2}a_{1}^{\dagger }a_{1}-\eta _{2}^{2}a_{2}^{\dagger
}a_{2}]\sum_{j}\sigma _{j}^{x}\text{,}
\end{equation}%
where the coupling constants between the two modes and individual spins are
\begin{equation}
\eta _{k}=\frac{\pi \mu _{0}S}{2\pi r\Phi _{0}}\sqrt{\frac{\hbar \omega _{k}%
}{Ll}}\text{.}  \label{eta}
\end{equation}%
For realistic parameters, $C_{\Sigma }=600$ aF, $C_{m}=30$ aF, $L_{0}=1$ cm,
$S_{0}=10\mu m^{2}$, $r=1\mu m$, $N=500$, $E_{J}=13$ GHz, we have $B=1.6$
GHz, $\omega _{2}\approx 120$ GHz, $\eta _{1}=\sqrt{3}\eta _{2}$, and $\eta
_{2}\approx 0.1$ \cite{Wang07}.

Thus, the total Hamiltonian is written as
\begin{align}
H =&E_J[(1-\frac{1}{2}\eta_1^2-\frac{1}{2}\eta_2^2)-\eta_1^2a_1^\dagger
a_1-\eta_2^2a_2^\dagger a_2]\sum_j \sigma _{j}^{x}  \notag \\
&+\frac{e^2C_m}{C_\Sigma^2}\sum_j\sigma _{j}^{z}\sigma
_{j+1}^{z}+\omega_1a_1^\dag a_1 +\omega_2a_2^\dag a_2 \text{.}
\label{originalH}
\end{align}

Furthermore, since there is no energy exchange between the fields and the
ICTF, the total Hamiltonian can be decomposed into invariant subspaces with
respect to the Fock state of the fields,
\begin{equation}
H=\sum_{m,n}H^{(m,n)}|m\rangle|n\rangle\langle n|\langle m|
\end{equation}
where
\begin{equation}
H^{(m,n)}=B\sum\limits_{j=1}^{N}(\lambda_{m,n} \sigma _{j}^{x}+\sigma
_{j}^{z}\sigma _{j+1}^{z})
\end{equation}
with $B=e^2C_m/C_\Sigma^2$ and $\lambda_{m,n}=E_J[1-(m+1/2)\eta_1^2-(n+1/2)%
\eta_2^2)]/B$.

Generally speaking, the Hamiltonian of ICTF $H_0$ is transformed into a
quadratic fermion form with Jordan-Wigner transformation \cite{Sachdev}
\begin{equation}
c_j=\exp(\pi i\sum\limits_{k=1}^{j-1}\sigma_k^z)\sigma_j^+ \text{.}
\end{equation}
Then, by introducing quasi-particle operator \cite{Pfeuty}
\begin{equation}
\gamma_k=\sum\limits_{j=1}^{N}\frac{e^{-ikj}}{\sqrt{N}}[\cos\frac{\theta_k}{2%
}c_j-i\sin\frac{\theta_k}{2}c_j^\dag] \text{,}
\end{equation}
with
\begin{equation}
\theta_k(\lambda)=\tan^{-1}\frac{\sin k}{\lambda-\cos k}\text{,}
\end{equation}
$H_0$ is diagonalized as
\begin{equation}
H_0=\sum\limits_{k}\varepsilon_k(\gamma_k^\dag\gamma_k-\frac{1}{2})
\end{equation}
with single particle spectrum being
\begin{eqnarray}
\varepsilon_{k}(\lambda)=2B\sqrt{1+\lambda^2-2\lambda\cos k}\text{.}
\end{eqnarray}
And the ground state $|G\rangle$ corresponds to no quasi-particle excitation
at all.

\section{Photon Bunching Effect}

Followed by a series of advances, i.e., resonance fluorescence, the
Hanbury-Brown-Twiss experiment \cite{Hanbury-Brown} reopen
philosophical debate about photons \cite{Knight} and set itself as
the milestone in the development of quantum optics. All these
experimental phenomena are associated with the correlation functions
of the field. Here, we consider it as the method to detect the QPT
since the two fields propagating in the TLR interact with the
quasi-spins respectively.

First of all, we define an operator
\begin{equation}
A=a_1+ia_2\text{.}
\end{equation}
The first order correlation function is written as $\langle
A^\dag(t)A\rangle$. Here, the bracket $\langle \cdots\rangle$
denotes average over the initial state, with the Ising chain in the
ground state $|G\rangle$ and the two fields being in arbitrary pure
states $\sum_m c_m |m\rangle$ and $\sum_n d_n |n\rangle$
respectively. Therefore,
\begin{align}
\langle A^\dag(t)A\rangle =&
\sum\limits_{m,n}|c_m|^2|d_n|^2(mr^{(m,n)}_{m-1,n}+nr^{(m,n)}_{m,n-1} )
\notag \\
&-ic_{m-1}^*c_m d_{n+1}^*d_n\sqrt{m(n+1)}r^{(m-1,n+1)}_{m-1,n}  \notag \\
&+ic_{m+1}^*c_m d_{n-1}^*d_n\sqrt{(m+1)n}r^{(m+1,n-1)}_{m,n-1}
\end{align}
where
\begin{equation}
r^{(m,n)}_{m^\prime,n^\prime}(t)=\langle
G|e^{iH^{(m,n)}t}e^{-iH^{(m^\prime,n^\prime)}t}|G\rangle  \label{rr}
\end{equation}
is the decoherence factor \cite{Quan06} which measures the overlap
of the ground state evolving under two different Hamiltonians.
Details about its calculation is presented in Appendix
\ref{app:appendix1}. In Ref.\cite{Wang07}, it was discovered that
for the same amount of environment dissipation the first order
correlation function of the single mode decreased more rapidly in
the vicinity of the QPT than in the other region.

Moreover, the second order correlation function $\langle A^\dag
A^\dag(t)A(t)A\rangle$ is analytically written as
\begin{widetext}
\begin{align}
&\langle A^\dag A^\dag(t)A(t)A\rangle= \notag \\
\sum\limits_{m,n}&|c_m|^2|d_n|^2[(m+n)(m+n-1)+mnr^{(m-1,n)}_{m,n-1}e^{-i(\omega_1-\omega_2)t}+mnr^{(m,n-1)}_{m-1,n}e^{i(\omega_1-\omega_2)t}] \notag \\
&+ic_{m+1}^*c_m d_{n-1}^*d_n\{(m+n-1)\sqrt{(m+1)n}+[m\sqrt{(m+1)n}r^{(m,n-1)}_{m-1,n}+(n-1)\sqrt{(m+1)n}r^{(m+1,n-2)}_{m,n-1}]e^{i(\omega_1-\omega_2)t}\} \notag \\
&-ic_{m-1}^*c_m d_{n+1}^*d_n\{(m+n-1)\sqrt{m(n+1)}+[(m-1)\sqrt{m(n+1)}r^{(m-2,n+1)}_{m-1,n}+n\sqrt{m(n+1)}r^{(m-1,n)}_{m,n-1}]e^{-i(\omega_1-\omega_2)t}\} \notag \\
&-c_{m+2}^*c_m d_{n-2}^*d_n
\sqrt{(m+2)(m+1)n(n-1)}r^{(m+1,n-2)}_{m,n-1}e^{i(\omega_1-\omega_2)t} \notag \\
&-c_{m-2}^*c_m d_{n+2}^*d_n
\sqrt{(m-1)m(n+1)(n+2)}r^{(m-2,n+1)}_{m-1,n}e^{-i(\omega_1-\omega_2)t}
\text{.}
\end{align}
\end{widetext}

Thus, for the fields initially in the state $(|0\rangle+|1\rangle)/%
\sqrt{2}$, it is straightforward to obtain
\begin{equation}
\langle A^\dag A^\dag(t)A(t)A\rangle=\frac{1}{2}[1+\text{Re}%
(r^{(1,0)}_{0,1}e^{i(\omega_1-\omega_2)t)})] \text{,}
\end{equation}
where Re$(x)$ means the real part of $x$.

As proven in Appendix \ref{app:appendix1}, in the vicinity of the QPT, the
square of the norm of $r^{(1,0)}_{0,1}(t)$ decreases more rapidly than
exponential, i.e,
\begin{equation}
|r^{(1,0)}_{0,1}(t)|^2\leq e^{-\gamma t^2} \text{,}
\end{equation}
where $\gamma=4B^2(\lambda_{1,0}-\lambda_{0,1})^2E(k_c)/(\lambda_{0,1}-1)^2$%
, $E(k_c)=4\pi^2N_c(N_c+1)(2N_c+1)/6N^2$ with $N_c$ being the nearest
integer to $Nk_c/2\pi$.

\begin{figure}[ptb]
\includegraphics[bb=81 260 485 564,width=7
cm]{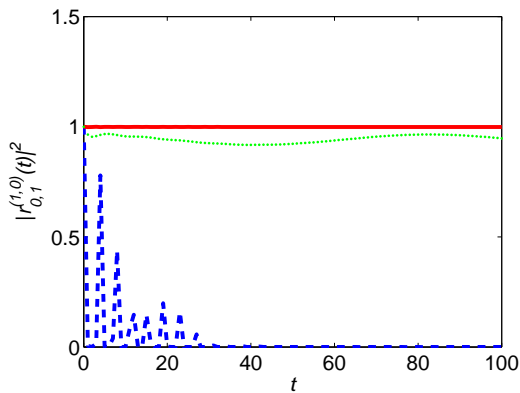}
\caption{ (color online) The decoherence factor $|r^{(1,0)}_{0,1}(t)|^2$ for
both fields in $(|0\rangle+|1\rangle)/\protect\sqrt{2}$ is plotted with $%
N=8000$. Blue dashed line for $\protect\lambda=1$, red solid line for $%
\protect\lambda=0.1$, and green dotted line for $\protect\lambda=2$. In all
figures, $t$ is in units of $1/B$.}
\label{r}
\end{figure}

It can be seen that there is a vanishing numerator $E(k_c)$ as $%
N\rightarrow\infty$. It is doubtful that the exponential decay of $%
|r^{(1,0)}_{0,1}(t)|^2$ can truly occur since the QPT takes place in the
thermodynamical limit. However, as the size of the ICTF gets larger, we can
adjust the parameter $\lambda_{0,1}$ closer to the critical point to make
the denominator $(\lambda_{0,1}-1)^2$ small enough. In that case, $\gamma$
stays as a constant and the $|r^{(1,0)}_{0,1}(t)|^2$ decreases exponentially
with time. For a real system, $N$ is finite for the demonstration of the
QPT. To test the validity of the above analysis, we resort to numerical
simulation. In Fig.\ref{r}, we plot the evolution of $|r^{(1,0)}_{0,1}(t)|^2$
according to Eq.(\ref{r1001}). It can be seen that despite some oscillations
$|r^{(1,0)}_{0,1}(t)|^2$ decays exceptionally at the critical point.

According to Ref.\cite{Scully}, the photon bunching and anti-bunching
effects are associated with the second order degree of coherence
\begin{equation}
g^{(2)}(t)=\frac{\langle A^\dag A^\dag(t)A(t)A\rangle}{\langle
A^\dag A\rangle \langle A^\dag(t) A(t)\rangle } \text{,}
\end{equation}
which is the normalized second order correlation function of the
fields. For the fields both in the state
$(|0\rangle+|1\rangle)/\sqrt{2}$, the second order degree of
coherence is simplified as
\begin{equation}
g^{(2)}(t)=\frac{1}{2}\frac{1+\text{Re}(r^{(1,0)}_{0,1}e^{i(\omega_1-%
\omega_2)t})}{1-\text{Im}(r^{(1,0)}_{0,1}e^{i(\omega_1-\omega_2)t})}
\end{equation}
with Im$(x)$ being the image part of $x$.

\begin{figure}[ptb]
\includegraphics[bb=24 208 552 637,width=7.5
cm]{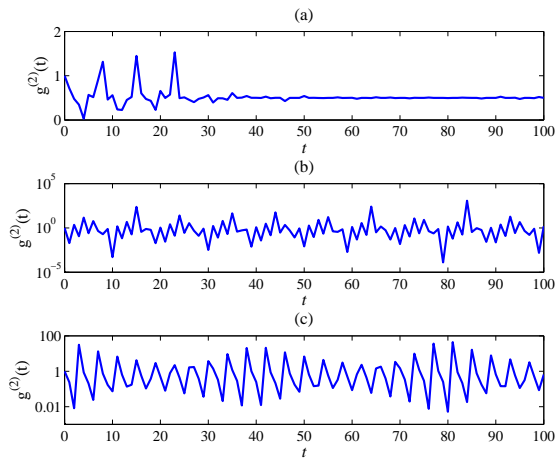} \caption{ (color online) The
second order degree of coherence $g^{(2)}(t)$ for $N=4000$ and
$(|0\rangle+|1\rangle)/\protect\sqrt{2}$ is plotted with (a) $\protect\lambda%
=1$, (b) $\protect\lambda=0.1$, (c) $\protect\lambda=2$. }
\label{g2lambda}
\end{figure}

\begin{figure}[ptb]
\includegraphics[bb=44 212 539 634,width=7.5
cm]{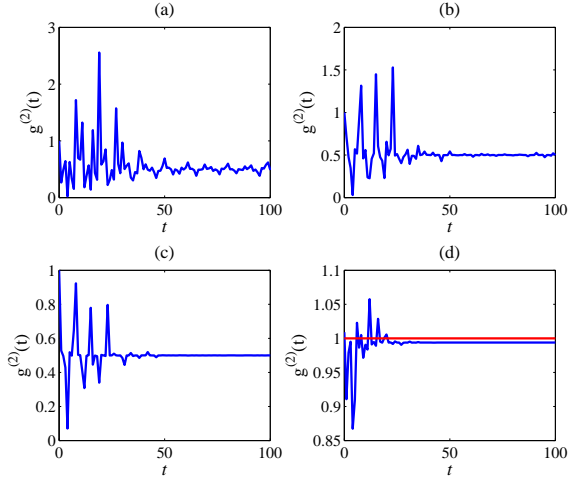} \caption{ (color
online) The second order degree of coherence $g^{(2)}(t)$ is plotted
at
the critical point for $(|0\rangle+|1\rangle)/\protect\sqrt{2}$ with (a) $%
N=2000$, (b) $N=4000$, (c) $N=8000$. For (d), both fields are in the
coherent states $|\protect\alpha\rangle$ with $\protect\alpha=1$ and
$N=8000$. Note that at the steady state $g^{(2)}(t)$ is a little
smaller than its original value $1$ as indicated by the red
horizontal line. } \label{g2N}
\end{figure}

Since the norm of the decoherence factor decreases exponentially at
the critical point, it is obvious that both the real and image parts
of $r^{(1,0)}_{0,1}e^{i(\omega_1-\omega_2)t}$ will vanish in that
limit. As a consequence, we expect the second order degree of
coherence to be less than unity in the steady state, i.e.,
$g^{(2)}(t)=1/2<g^{(2)}(0)=1 $. Generally speaking, classical fields
such as thermal light and coherent light, prefer to distribute
themselves in bunches rather than at random. They exhibit less
correlation for time longer than the correlation time. This is so
called bunching effect \cite{Scully}. On the contrary, in certain
quantum optical systems, fewer quantum photons are detected close
together than further apart. And the photon antibunching observed in
fluorescent light from a two-level atom \cite{Knight} is of such
kind. Here, since the two fields involved are two independent modes,
we expect photons to be neither bunching nor antibunching,
regardless of quantum mechanical fields or classical fields.
However, as shown in Fig.\ref{g2lambda}, when the Ising
chain is at the critical point, the two independent fields initially in $%
(|0\rangle+|1\rangle)/\sqrt{2}$ display the photon bunching effect.
Further witness is also demonstrated in Fig.\ref{g2N}(a-c). It can
also be proven that $g^{(2)}(t)<g^{(2)}(0)$ for both fields in the
coherent state $|\alpha\rangle$ which is not shown here. In
Fig.\ref{g2N}(d), we plot the time evolution of the second order of
coherence for this case. Here, we remark that the two initially
independent quantum fields display the classical effect due to their
common interaction with the quantum critical system. As illustrated
in Eq.(\ref{rr}), two initially identical states evolve under two
slightly different Hamiltonians. Although the difference between
these Hamiltonians are tiny, their evolution trajectories are quite
distinct in the vicinity of the QPT. Thus, this slight difference
leads to the exponential decay of their decoherence factor. It can
be understood as a signature of quantum chaos \cite{Quan06}.

Furthermore, for the parameters mentioned after Eq.(\ref{eta}) and
$N_c=N/10$, both the real and imaginary parts of
$r^{(1,0)}_{0,1}e^{i(\omega_1-\omega_2)t}$ decay with a rate of the
order $\sqrt{\gamma}\simeq2.5$ GHz. Since the dissipation rate of
the first excitation mode is about $6.3$ MHz \cite{Wallraff}, we can
neglect the influence due to the dissipation of TLR.

\section{Conclusion and Remark}

To conclude, we have explored the possibility to probe quantum
criticality in the ICTF by detecting the higher order quantum
coherence of the two modes of cavity fields coupled to the spins. We
suggest a physical implementation of this theoretical scheme based
on a circuit QED system where the capacitively coupled CPBs are
coupled to the TLR. Situated at the antinodes of both modes
propagating in the TLR, CPBs are only coupled to the magnetic
fields. In a heuristic way, we show the decoherence factor decays
exponentially with time in the vicinity of the critical point. The
second order of coherence is smaller than one at the steady state.
Thus, the two initially independent modes demonstrate photon
bunching effect. This can serve as a witness of the QPT.

On the other hand, we have not investigated decoherence originated
from the dissipation of the CPBs. We notice that in a recent work
\cite{Hoyos}, the QPT in the dissipative random transverse-field
Ising chain was investigated. It was discovered that the quantum
critical point was ruined by the interplay between quantum
fluctuations and Ohmic dissipation. Further exploration may be done
when such kind of effect is considered.

\section*{Acknowledgement}

One (A. Q.) of the authors thanks W. Y. Huo for warm discussions.
This work is partially supported by the National Fundamental
Research Program Grant No. 2006CB921106, China National Natural
Science Foundation Grant Nos. 10325521, 60635040. Y.D.W. is
supported by the ECIST-FET project EuroSQUIP, the Swiss SNF, and the
NCCR Nanoscience.

\appendix

\section{Decoherence Factor}

\label{app:appendix1}

Following the method introduced in Ref.\cite{Wang07}, the decoherence factor
$r^{(m,n)}_{m^\prime,n^\prime}$ can be calculated in the following way.

By introducing the spin-1 pseudospin operators \cite{Anderson}
\begin{align}
s_{xk} &= i(\gamma_{-k}\gamma_k + \gamma_{-k}^\dag\gamma_k^\dag) \text{,}
\notag \\
s_{yk} &= \gamma_{-k}^\dag\gamma_k^\dag - \gamma_{-k}\gamma_k \text{,}
\notag \\
s_{zk} &= \gamma_k^\dag\gamma_k + \gamma_{-k}^\dag\gamma_{-k}-1 \text{,}
\end{align}
the Hamiltonian $H_0$ can also be rewritten as
\begin{equation}
H_0=\sum\limits_{k>0}\varepsilon_k s_{zk} \text{.}
\end{equation}

Because there is no energy exchange between the two modes and the qubit
array, the total Hamiltonian can be decomposed into invariant subspaces with
respect to the Fock state of the fields, i.e., $H=\sum_{m,n}H^{(m,n)}|m%
\rangle|n\rangle\langle n|\langle m|$, where
\begin{equation}
H^{(m,n)}=B\sum\limits_{j=1}^{N}(\lambda_{m,n} \sigma _{j}^{x}+\sigma
_{j}^{z}\sigma _{j+1}^{z}) \text{.}
\end{equation}
With the pseudospin operators, we can also diagonalize the Hamiltonian as
\begin{align}
H^{(m,n)}=\sum\limits_{k>0}\varepsilon_k^{(m,n)}s^{(m,n)}_{zk} \text{,}
\end{align}
where $s^{(m,n)}_{zk}=s_{zk}\cos2\alpha^{(m,n)}_k+s_{xk}\sin2\alpha^{(m,n)}_k
$ , with $2\alpha^{(m,n)}_k=\theta^{(m,n)}_k-\theta_k$, $%
\varepsilon^{(m,n)}_k=\varepsilon_k(\lambda_{m,n})$, $\theta^{(m,n)}_k=%
\theta_k(\lambda_{m,n})$.

Therefore, the ground state of $H_0$ is the product state of all pseudospins
down $|-\rangle_k$,
\begin{align}
|G\rangle&=\prod\limits_{k>0}^\bigotimes|-\rangle_k  \notag \\
&=\prod\limits_{k>0}^\bigotimes
(\cos\alpha^{(m,n)}_k|-\rangle^{(m,n)}_k+\sin\alpha^{(m,n)}_k|+%
\rangle^{(m,n)}_k)
\end{align}
with $|\pm\rangle^{(m,n)}_k$ being the eigen states of $s^{(m,n)}_{zk}$.

Since
\begin{align}
_k ^{(m,n)}\langle\pm|\pm\rangle _{k^\prime}^{(m^\prime,n^\prime)}
&=\delta_{kk^\prime}\cos(\alpha_k^{(m,n)}-\alpha_k^{(m^\prime,n^\prime)})
\text{,}  \notag \\
_k ^{(m,n)}\langle+|-\rangle _{k^\prime}^{(m^\prime,n^\prime)}
&=\delta_{kk^\prime}\sin(\alpha_k^{(m,n)}-\alpha_k^{(m^\prime, n^\prime)})
\text{,}  \notag \\
_k ^{(m,n)}\langle-|+\rangle _{k^\prime}^{(m^\prime,n^\prime)}
&=\delta_{kk^\prime}\sin(\alpha_k^{(m^\prime, n^\prime)}-\alpha_k^{(m,n)})
\text{,}
\end{align}
we have
\begin{align}
r^{(m,n)}_{m^\prime,n^\prime}(t)=\prod\limits_{k>0}\sum\limits_{a_k,b_k=%
\pm}C_{a_k,b_k,k}^{(m,n,m^\prime,n^\prime)}e^{i(a_k\varepsilon_k^{(m,n)}+b_k%
\varepsilon_k^{(m^\prime,n^\prime)})t}
\end{align}
with
\begin{align}
C_{+,+,k}^{(m,n,m^\prime,n^\prime)}
&=\sin\alpha_k^{(m,n)}\cos\alpha_k^{(m^\prime,
n^\prime)}\sin(\alpha_k^{(m,n)}-\alpha_k^{(m^\prime,n^\prime)}) \text{,}
\notag \\
C_{-,-,k}^{(m,n,m^\prime,n^\prime)}
&=\cos\alpha_k^{(m,n)}\sin\alpha_k^{(m^\prime,
n^\prime)}\sin(\alpha_k^{(m^\prime,n^\prime)}-\alpha_k^{(m,n)}) \text{,}
\notag \\
C_{+,-,k}^{(m,n,m^\prime,n^\prime)}
&=\sin\alpha_k^{(m,n)}\sin\alpha_k^{(m^\prime,
n^\prime)}\cos(\alpha_k^{(m,n)}-\alpha_k^{(m^\prime,n^\prime)}) \text{,}
\notag \\
C_{-,+,k}^{(m,n,m^\prime,n^\prime)}
&=\cos\alpha_k^{(m,n)}\cos\alpha_k^{(m^\prime,
n^\prime)}\cos(\alpha_k^{(m,n)}-\alpha_k^{(m^\prime,n^\prime)}) \text{.}
\notag \\
\end{align}

For heuristic analysis, we obtain the short time behavior of $%
|r^{(1,0)}_{0,1}(t)|^2$ at the critical point.
\begin{widetext}
\begin{align}
|r^{(1,0)}_{0,1}(t)|^2 = \prod\limits_{k>0} & F_k \notag \\
=\prod\limits_{k>0} & [\sin^2(\alpha_k^{(1,0)}-\alpha_k^{(0,1)})\cos(\varepsilon_k^{(1,0)}+\varepsilon_k^{(0,1)})t+\cos^2(\alpha_k^{(1,0)}-\alpha_k^{(0,1)})\cos(\varepsilon_k^{(1,0)}-\varepsilon_k^{(0,1)})t]^2 \notag \\
&+[\sin(\alpha_k^{(1,0)}+\alpha_k^{(0,1)})\sin(\alpha_k^{(1,0)}-\alpha_k^{(0,1)})\cos(\varepsilon_k^{(1,0)}+\varepsilon_k^{(0,1)})t
\notag
\\&-\cos(\alpha_k^{(1,0)}+\alpha_k^{(0,1)})\cos(\alpha_k^{(1,0)}-\alpha_k^{(0,1)})\sin(\varepsilon_k^{(1,0)}+\varepsilon_k^{(0,1)})t]^2
\label{r1001}
\end{align}
\end{widetext}
Since all factors $F_k$ of $|r^{(1,0)}_{0,1}(t)|^2$ have a norm less than
unity, we may expect the $|r^{(1,0)}_{0,1}(t)|^2$ to vanish under certain
conditions. Here, we set a cutoff frequency $k_c$ and hence we have $%
|r^{(1,0)}_{0,1}(t)|^2\leq\prod\limits_{k>0}^{k_c}F_k$. For small $k$, we
have
\begin{align}
\varepsilon_k^{(1,0)}&\approx2B|1-\lambda_{1,0}| \text{,}  \notag \\
\varepsilon_k^{(0,1)}&\approx2B|1-\lambda_{0,1}| \text{,}  \notag \\
\theta_k(\lambda)&\approx\frac{k}{\lambda-1} \text{,}  \notag \\
\alpha_k^{(1,0)}&\approx\frac{1}{2}(\frac{k}{\lambda_{1,0}-1}-\frac{k}{%
\lambda_{0,0}-1}) \text{,}  \notag \\
\alpha_k^{(0,1)}&\approx\frac{1}{2}(\frac{k}{\lambda_{0,1}-1}-\frac{k}{%
\lambda_{0,0}-1}) \text{.}  \notag
\end{align}
To the second order of $\alpha_k^{(1,0)}\pm\alpha_k^{(0,1)}$, we obtain
\begin{align}
&|r^{(1,0)}_{0,1}(t)|^2  \notag \\
&\leq\prod\limits_{k>0}^{k_c}[1-2(\alpha_k^{(1,0)}-\alpha_k^{(0,1)})^2]%
\cos^2(\varepsilon_k^{(1,0)}-\varepsilon_k^{(0,1)})t  \notag \\
&+2(\alpha_k^{(1,0)}-\alpha_k^{(0,1)})^2\cos(\varepsilon_k^{(1,0)}+%
\varepsilon_k^{(0,1)})t\cos(\varepsilon_k^{(1,0)}-\varepsilon_k^{(0,1)})t
\notag \\
&+[1-(\alpha_k^{(1,0)})^2-(\alpha_k^{(0,1)})^2]\sin^2(\varepsilon_k^{(1,0)}-%
\varepsilon_k^{(0,1)})t  \notag \\
&-2[(\alpha_k^{(1,0)})^2-(\alpha_k^{(0,1)})^2]\sin(\varepsilon_k^{(1,0)}-%
\varepsilon_k^{(0,1)})t  \notag \\
&\times\sin(\varepsilon_k^{(1,0)}+\varepsilon_k^{(0,1)})t \text{.}  \notag
\end{align}
Since
\begin{equation}
0\approx\varepsilon_k^{(1,0)}-\varepsilon_k^{(0,1)}\ll\varepsilon_k^{(1,0)}+%
\varepsilon_k^{(0,1)}\approx2\varepsilon_k^{(1,0)} \text{,}  \notag
\end{equation}
we focus on the short time behavior and therefore
\begin{align}
&|r^{(1,0)}_{0,1}(t)|^2  \notag \\
&\leq\prod\limits_{k>0}^{k_c}[1-2(\alpha_k^{(1,0)}-\alpha_k^{(0,1)})^2]+2(%
\alpha_k^{(1,0)}-\alpha_k^{(0,1)})^2\cos(2\varepsilon_k^{(1,0)}t)  \notag \\
&=\prod\limits_{k>0}^{k_c}1-2(\alpha_k^{(1,0)}-\alpha_k^{(0,1)})^2\sin^2(%
\varepsilon_k^{(1,0)}t)  \notag \\
&=\prod\limits_{k>0}^{k_c}1-2\frac{k^2(\lambda_{1,0}-\lambda_{0,1})^2}{%
(\lambda_{1,0}-1)^2(\lambda_{0,1}-1)^2}\sin^2(2Bt|\lambda_{1,0}-1|) \text{.}
\notag
\end{align}
As $\lambda_{1,0}\rightarrow1$, we have
\begin{equation}
|r^{(1,0)}_{0,1}(t)|^2\leq e^{-\gamma t^2} \text{,}
\end{equation}
where $\gamma=4B^2(\lambda_{1,0}-\lambda_{0,1})^2E(k_c)/(\lambda_{0,1}-1)^2$%
, $E(k_c)=4\pi^2N_c(N_c+1)(2N_c+1)/6N^2$ with $N_c$ being the nearest
integer to $Nk_c/2\pi$.

\end{document}